\documentclass[12p]{article}

\usepackage[T1]{fontenc}
\usepackage[latin1]{inputenc}
\usepackage{eurosym}

\usepackage{latexsym,amssymb,amsmath,amscd,amscd,amsthm,amsxtra}
\usepackage[dvips]{graphicx}

\newtheorem{thm}{Theorem}[section]
\newtheorem{lem}[thm]{Lemma}

\newtheorem{pro}[thm]{Proposition}
\newtheorem{ex}[thm]{Example}

\newtheorem{defi}[thm]{Definition}

\def\gpd{\,\lower1pt\hbox{$\longrightarrow$}\hskip-.24in\raise2pt
         \hbox{$\longrightarrow$}\,}
\def\qed{\hfill ~\vrule height6pt width6pt depth0pt}
\newcommand{\CC}{\mathbb{C}}

\begin{document}
\title{Poisson Cohomology of Del Pezzo surfaces }
\author{Wei Hong\\ 
Department of Mathematics\\
Pennsylvania State University \\
University Park, PA 16802, USA\\
\textsf{hong\textunderscore w@math.psu.edu} \\
 \and
Ping Xu\thanks{Research partially supported by NSF grants DMS-0605725 and DMS-0801129, 
and NSFC grant 10911120391/A0109.} \\
Unit\'e de Recherche en Math\'ematiques\\
Universit\'e du Luxembourg\\ 
1359 Luxembourg\\
Grand-Duch\'e de Luxembourg\\ \\
Department of Mathematics\\
Pennsylvania State University \\
University Park, PA 16802, USA\\
\textsf{ping@math.psu.edu}
}

\maketitle

\begin{abstract} In this paper,
 we compute the Poisson cohomology groups for any Poisson Del Pezzo surface.
\end{abstract}


\section{Introduction}

Recently, there has been increasing interest on holomorphic
Poisson structures due to their close connection
to mirror symmetry \cite{K2}. Holomorphic
Poisson structures are also known to be connected
to generalized complex structures \cite{Hitchin:03, Cal}
and biHamiltonian systems \cite{LSX}. To a holomorphic Poisson manifold
 $(X, \pi)$, Poisson cohomology groups are important invariants.
If $X$ is  compact, then Poisson cohomology groups are
known  \cite{Stienon, CSX} to be finite dimensional.

A Del Pezzo surface is a two-dimensional Fano variety, i.e. an
algebraic surface with ample anti-canonical divisor class.
For the  dimension reason, any holomorphic bivector field
on a Del Pezzo surface is automatically a holomorphic Poisson
tensor. Therefore, holomorphic Poisson structures on a del Pezzo
 surface should provide us with many interesting examples of
holomorphic Poisson structures on compact manifolds.
 It is natural to ask what the corresponding
 Poisson cohomology groups are. In this paper, we first
give an explicit description of all holomorphic bivector fields
on a given del Pezzo surface $X$.  The result  is classical
and can be found scattered, such as in  \cite{kod}. For the
reader's convenience, in Section 2,  we recall all the results, which will
also be important for our computations.  Our approach relies on
the classification of del Pezzo surfaces: they are 
$\mathbb{CP}^{1}\times \mathbb{CP}^{1}$, $\mathbb{CP}^{2}$
or  $\mathbb{CP}^{2}$ blow up at $r$ ($1\leq r\leq 8$) generic
points. For the computation of  Poisson cohomology groups,
we use the Dolbeault resolution, i.e., the double
complex as in  \cite{LSX}. Since the Poisson cohomology groups
are over $\CC$ and are finite dimensional, our  approach
essentially reduces to the computation of
their dimensions.

Various special cases of holomorphic Poisson structures
considered here have appeared in literature. For instance,
holomorphic Poisson structures on $\mathbb{CP}^{1}\times \mathbb{CP}^{1}$
were studied by Duistermaat \cite{Duistermaat}, and
a certain holomorphic Poisson structure on $\mathbb{CP}^{2}$
appeared naturally in the work of Evens-Lu \cite{EL} while studying
the variety of Lagrangian subalgebras.
Noncommutative Del Pezzo surfaces were studied by several
authors, for instance, \cite{AKO, Bergh1, Bergh2, EG}.
It would be interesting to 
 explore  the relationship  to these works,  which will be discussed somewhere else.

\paragraph{Acknowledgments}
We would like to thank Pennsylvania State University (Hong)  and Peking
University (Xu) for their hospitality while work on this project
was being done.
We also wish to thank Sam Evens, Jiang-hua Lu, Tony Pantev,  Mathieu Stienon, and
Dmitry Tarmarkin for their useful discussions and comments.
Hong's research was 
 partially supported by CSC grant and NSF
grant DMS-0605725, and  
Xu's research was partially supported 
 by NSF grants DMS-0605725 and DMS-0801129, NSA grant H98230-10-1-0195
 and NSFC  grant 10911120391/A0109.

\section{Preliminary}
In this section,  we recall some basic results about Del Pezzo
surfaces. A Del Pezzo surface is a two dimensional Fano variety, i.e. an
algebraic surface with ample anti-canonical divisor class. The following is a classical result:

\begin{thm}
 \cite{Manin} A Del Pezzo surface is isomorphic to one of the following cases:
 \begin{enumerate}
\item $\mathbb{CP}^{1}\times \mathbb{CP}^{1}$
\item $\mathbb{CP}^{2}$
\item $\mathbb{CP}^{2}$ blow up at r ($1\leq r\leq 8$) generic
points $p_{i}$ $(i=1,2,...r)$.
\end{enumerate}
\end{thm}
Recall that $p_{1},...,p_{r}$ are said to be generic if the following
conditions are satisfied: none of any three are colinear and
none of any six points are on the same conic. We will denote $B_{r}$
as the blow up $\mathbb{CP}^{2}$ at r ($1\leq r\leq 8$) generic
points. By $\rho$,  we denote the projection map from $B_{r}$ to
$\mathbb{CP}^{2}$.

Most of the result below are classical and can be found in
$\cite{kod}$.
\begin{pro}
\begin{enumerate}
\item If $X=\mathbb{CP}^{1}\times \mathbb{CP}^{1}$, then
\begin{equation}\label{cp11}
 \dim H^{0}(X,\wedge^{2}T_{X})=9,\quad \dim H^{0}(X,T_{X})=6.
 \end{equation}
 \item If $X=\mathbb{CP}^{2}$, then
\begin{equation}\label{cp111}
 \dim H^{0}(X,\wedge^{2}T_{X})=10,\quad \dim H^{0}(X,T_{X})=8.
 \end{equation}
 \end{enumerate}
\end{pro}
\proof
\begin{enumerate}
 \item  $X=\mathbb{CP}^{1}\times \mathbb{CP}^{1}$ has four open covers
$$U_{1}=\{([z_{0},z_{1}],[w_{0},w_{1}])|z_{0}\neq0,w_{0}\neq0\},\qquad
U_{2}=\{([z_{0},z_{1}],[w_{0},w_{1}])|z_{1}\neq0,w_{0}\neq0\},$$
$$U_{3}=\{([z_{0},z_{1}],[w_{0},w_{1}])|z_{0}\neq0,w_{1}\neq0\},\qquad
U_{4}=\{([z_{0},z_{1}],[w_{0},w_{1}])|z_{1}\neq0,w_{1}\neq0\}.$$ Let
$x=\frac{z_{1}}{z_{0}}$, $w=\frac{w_{1}}{w_{0}}$,
$x^{\prime}=\frac{z_{0}}{z_{1}}=\frac{1}{x}$,
$w^{\prime}=\frac{w_{0}}{w_{1}}=\frac{1}{w}$ be their corresponding
affine coordinates. It is clear that
$$\frac{\partial}{\partial x}
=-(x^{\prime})^{2}\frac{\partial}{\partial x^{\prime}}\qquad
\frac{\partial}{\partial w}
=-(w^{\prime})^{2}\frac{\partial}{\partial w^{\prime}}$$ Assume that
$v$ is a holomorphic vector field. Write
$$ v=f(x,w)\frac{\partial}{\partial x}+g(x,w)\frac{\partial}{\partial w}$$
on $U_{1}$, where $f(x,w)$ and $g(x,w)$ are holomorphic functions on
$U_{1}$. On $U_{1}\cap U_{2}$, by coordinate change
$x^{\prime}=\frac{1}{x}$, $v$ takes the following form in coordinate
$(x^{\prime},w)$
$$ v=-f(\frac{1}{x^{\prime}},w)(x^{\prime})^{2}
\frac{\partial}{\partial x^{\prime}}+
g(\frac{1}{x^{\prime}},w)\frac{\partial}{\partial w}
$$
Since $v$ is holomorphic on $U_{2}$, it follows that the power
series of $f(x,w)$ about $x$ is up to at most $x^{2}$, and
$g(x,w)$ is independent of $x$.
Similarly, $f(x,w)$ is independent of $w$, and the power series of
$g(x,w)$ about $w$ is up to at most $w^{2}$.
 Hence on $U_{1}$, $v$ can be written as
\begin{equation}\label{cp1v}
v=(b_{1}+b_{2}x+b_{3}x^{2})\frac{\partial}{\partial x}+
(b_{4}+b_{5}w+b_{6}w^{2})\frac{\partial}{\partial w},
\end{equation}
where $b_{i}\in\mathbb{C}$, $i=1,2, \cdots, 6$, are constants.

 Let $\pi$ be a holomorphic bivector field. By a similar argument, we see that,
on $U_{1}$, $\pi$ can be written as
\begin{equation}\label{cp1b}
\pi=(a_{1}+a_{2}x+a_{3}x^{2}+a_{4}w+a_{5}xw+a_{6}x^{2}w+
a_{7}w^{2}+a_{8}xw^{2}+a_{9}x^{2}w^{2})\frac{\partial}{\partial x}\wedge\frac{\partial}{\partial w},
\end{equation}
where $a_{i}\in\mathbb{C}$, $i=1,2,...9$, are constants. As a
consequence, we have
 \begin{equation}
 \dim H^{0}(X,\wedge^{2}T_{X})=9,\quad \dim H^{0}(X,T_{X})=6.
 \end{equation}

 \item  Let $U_{i}=\{[z_{0},z_{1},z_{2}]|z_{i}\neq 0\},
i=0,1,2,$ be an  open cover of $\mathbb{CP}^{2}$. Let
$x=\frac{z_{1}}{z_{0}}$,
  $w=\frac{z_{2}}{z_{0}}$ be the affine coordinates on $U_{0}$.
  If $\pi$ is a holomorphic bivector field on X, by a similar argument,
  one shows that $\pi$ can be written as
 \begin{equation}\label{cp2b}
 \pi=(a_{1}+a_{2}x+a_{3}w+a_{4}x^{2}+a_{5}xw
 +a_{6}w^{2}+a_{7}x^{3}+a_{8}x^{2}w+a_{9}xw^{2}+a_{10}w^{3})
 \frac{\partial}{\partial x}\wedge\frac{\partial}{\partial w}.
  \end{equation}
 Similarly, if $v$ is a holomorphic vector field on $\mathbb{CP}^{2}$,
 then $v$ can be written as (\cite{kod},~~p223)
 \begin{equation}\label{cp2v}
 v=(b_{1}+b_{2}x+b_{3}w+b_{7}x^{2}+b_{8}xw)
 \frac{\partial}{\partial x}+
 (b_{4}+b_{5}x+b_{6}w+b_{7}xw+b_{8}w^{2})
 \frac{\partial}{\partial w}.
 \end{equation}
 As a consequence, we have
 \begin{equation}\label{cp2}
 \dim H^{0}(X,\wedge^{2}T_{X})=10,\quad \dim H^{0}(X,T_{X})=8.
 \end{equation}
 \end{enumerate}
 \qed

 Now we consider the case $B_{r}$, i.e $\mathbb{CP}^{2}$ with blow up at
 $r$ generic  points  $p_{i},i=1,2,...r$.
 Let $\rho: B_{r}\rightarrow \mathbb{CP}^{2}$ be the projection map.
 Write $$V^{1}_{r}= \{v|v~~~\text{is a holomorphic vector field on}~~~\mathbb{CP}^{2}~~~
 \text{such that}~~~ v(p_{i})=0, i=1,2,..r\}$$
 $$V^{2}_{r}= \{\pi|\pi~~~\text{is a holomorphic bivector field on}~~~\mathbb{CP}^{2}~~~
  \text{such that}~~~ \pi(p_{i})=0, i=1,2,...r\}.$$

The following can be found in (\cite{kod},~p225).
 \begin{lem}\label{kodb}
 \cite{kod} $\rho_{*}$ induces an isomorphism from $H^{0}(B_{r},T_{B_{r}})$ to $V^{1}_{r}$
 and an isomorphism from $H^{0}(B_{r},\wedge^{2}T_{B_{r}})$ to $V^{2}_{r}$.
 \end{lem}

\begin{lem}
 \label{ag}(\cite{G-H}, p481)
 Eight points $p_{i}\in \mathbb{CP}^{2}(i=1,2,...8)$ fail to impose independent conditions on cubics only
 when all eight points lie on a conic curve, or five of them   are colinear.
 \end{lem}

Assume that $\{f_{1}, \cdots ,f_{10}\}$
is a basis of the space of  cubic polynomials on $\mathbb{CP}^{2}$.
 Recall that $p_{i}\in \mathbb{CP}^{2}(i=1,2,...8)$ satisfy the independent conditions on cubics means
 that the $10\times8$ matrix $(f_{j}(p_{i}))$ $(1\leq i\leq8,1\leq j\leq10)$ is of full rank.

\begin{pro}
For $X=B_{r}$, we have
\begin{eqnarray}
 \label{bhv}& &\dim H^{0}(X,T_{X})=
 \begin{cases}
 8-2r,& 1\leq r\leq3\\
 0,& r\geq4
 \end{cases}\\
 \label{bhb}& &\dim H^{0}(X,\wedge^{2}T_{X})=10-r.
 \end{eqnarray}
\end{pro}
\proof
By Lemma (\ref{kodb}),$$ H^{0}(B_{r},T_{B_{r}})\cong V^{1}_{r}=\{v|v~~~
\text{is a holomorphic vector field on}~~~\mathbb{CP}^{2}~~~\text{such that}~~~ v(p_{i})=0,i=1,2,...r\}.$$
If $r\geq 4$, we can choose the coordinates of
$p_{1},p_{2},p_{3},p_{4}$ as $[1,0,0], [1,1,0], [1,0,1],[1,1,1]$
 since any three of them are not
colinear. By a simple computation using Equation (\ref{cp2v}), it
 is easy to see that $v$ must be equal to zero. \\
If $1\leq r\leq 3$, by the same argument, we have
 $$\dim H^{0}(B_{r},T_{B_{r}})=8-2r. $$
 For Equation (\ref{bhb}), by Lemma (\ref{kodb}),
$$ H^{0}(B_{r},\wedge^{2}T_{B_{r}})\cong V^{2}_{r}=
\{\pi|\pi~~~\text{is a holomorphic bivector field on}~~~\mathbb{CP}^{2}~~~\text{such that}~~~\pi(p_{i})=0,i=1,2,...r\}.$$

Without loss of generality, we assume that all
points $p_{i}\in U_{0}=\{[z_{0},z_{1},z_{2}]|z_{0}\neq 0\}$.
According to Lemma \ref{ag}, $\{ p_{i}\}$ should satisfy the independent
condition on cubics. Therefore they also satisfy the independent condition
under the affine coordinates. By Equation (\ref{cp2b}), we have
$$\dim H^{0}(B_{r},\wedge^{2}T_{B_{r}})=10-r.$$
\qed

  \begin{thm}\label{del}
   For $X=\mathbb{CP}^{1}\times \mathbb{CP}^{1}$, $\mathbb{CP}^{2}$, or $B_{r}(1\leq r\leq4)$,  we have
 \begin{equation}
 H^{i}(X,\wedge^{j}T_{X})=0,~~~for~~i>0.
 \end{equation}
 For $X=B_{r} (5\leq r\leq8)$, we have
\begin{equation}
 H^{i}(X,\wedge^{j}T_{X})=
 \begin{cases}
 0 & i>0~~~and~~~(i,j)\neq(1,1)\\
 2r-8 & (i,j)=(1,1)
 \end{cases}
 \end{equation}
  \end{thm}
 \proof
  \begin{enumerate}
   \item Consider $X=\mathbb{CP}^{1}\times \mathbb{CP}^{1}$.\\
    By $\rho_{1}$ and $\rho_{2}$, we denote the projection maps from 
   $\mathbb{CP}^{1}\times \mathbb{CP}^{1}$ to the first and the second component, respectively. Let
   $L_{i}=\rho_{i}^{*}T_{\mathbb{CP}^{1}}$, $i=1,2,$ be the pull back line bundles on X.
   Then $T_{X}\cong L_{1}\oplus L_{2}$. Hence
   \begin{equation*}
   H^{i}(X,T_{X})\cong H^{i}(X,L_{1}\oplus L_{2})\cong H^{i}(X,L_{1})\oplus H^{i}(X,L_{2}).
   \end{equation*}
   Since $\wedge^{2} T_{X}\cong L_{1}\otimes L_{2}$, it thus follows that
   $$H^{i}(X,L_{1})\cong H^{i}(X,\mathcal{K}_{X}\otimes L_{1}^{2}\otimes L_{2}),$$
   where $\mathcal{K}_{X}$ is the canonical line bundle over $X$.
   Since $L_{1}^{2}\otimes L_{2}$ is a positive line bundle, by Kodaira vanishing theorem, we have
   $H^{i}(X,L_{1})=0,\forall i>0$. Similarly $H^{i}(X,L_{2})=0, \forall i>0$.
   Hence
   \begin{equation}
   H^{i}(X,T_{X})=0,\forall~~~i>0.
   \end{equation}
   Similarly, we have
   \begin{equation}
    H^{i}(X,\wedge^{2}T_{X})\cong H^{i}(X,L_{1}\otimes L_{2})\cong H^{i}(X,\mathcal{K}_{X}
    \otimes L_{1}^{2}\otimes L_{2}^{2}),
   \end{equation}
   and
   \begin{equation}
    H^{i}(X,\mathcal{O}_{X})=H^{i}(X,\mathcal{K}_{X}\otimes L_{1}\otimes L_{2}).
   \end{equation}
    Since both $L_{1}^{2}\otimes L_{2}^{2}$ and $L_{1}\otimes L_{2}$ are positive line bundles,
     by Kodaira vanishing theorem,
    we obtain
     \begin{equation}
     H^{i}(X,\wedge^{2}T_{X})=0, ~~~\forall i>0, \ \ \ \mbox{and}
     \end{equation}
     \begin{equation}
     H^{i}(X,\mathcal{O}_{X})=0, ~~~\forall i>0.
     \end{equation}

 \item Let $X=\mathbb{CP}^{2}$. Then
 \begin{equation}
H^{i}(X,\mathcal{O}_{X})=H^{0,i}(X)=0,\quad \forall  i>0.
\end{equation}
The Euler sequence
$$ 0\rightarrow \mathcal{O}_{X}\rightarrow \overset{3}{\underset{i=1}{\oplus}} \mathcal{O}(1)
\rightarrow T_{X}\rightarrow 0$$
induces a long exact sequence on the level of cohomology:
$$\rightarrow H^{i}(X,\mathcal{O}_{X})\rightarrow H^{i}(X,
\overset{3}{\underset{i=1}{\oplus}}\mathcal{O}(1))\rightarrow
H^{i}(X,T_{X}) \rightarrow H^{i+1}(X,\mathcal{O}_{X})\rightarrow $$
 If $i>0$, $H^{i}(X,\mathcal{O}_{X})=0$, and therefore $H^{i}(X,
 \overset{3}{\underset{i=1}{\oplus}} \mathcal{O}(1))\cong H^{i}(X,T_{X})$.
 Since $$H^{i}(X,\overset{3}{\underset{i=1}{\oplus}}\mathcal{O}(1))=
 \overset{3}{\underset{i=1}{\oplus}} H^{i}(X,\mathcal{O}(1))=0,$$
  we have
  \begin{equation}
  H^{i}(X,T_{X})=0,~~~\forall ~~~i>0.
  \end{equation}
  On the other hand, since $\wedge^{2}T_{X}\cong \mathcal{O}(3)$, we have $$H^{i}(X,\wedge^{2}T_{X})=
 H^{i}(X,\mathcal{O}(3))= H^{i}(X,\mathcal{K}_{X}\otimes\mathcal{O}(6)).$$
  Since $\mathcal{O}(6)$ is a positive line bundle, by Kodaira vanishing theorem we have
  \begin{equation}
  H^{i}(X,\wedge^{2}T_{X})=0,~~~\forall i>0.
  \end{equation}

 \item Let $X=B_{r}(1\leq r\leq4)$.\\
 Since $X$ is an algebraic varity, we have
 $$H^{i}(X,\mathcal{O}_{X})\cong H^{0,i}(X)\cong H^{i,0}(X)\cong H^{0}(X,\Omega^{i}_{X}).$$
   According to (\cite{kod},p225),
 $H^{0}(X,\Omega^{i}_{X})=0$ if $i>0$. Hence $H^{i}(X,\mathcal{O}_{X})=0$,  $\forall i>0$.
 According to (\cite{kod}, p225-226), $H^{1}(X,T_{X})=0$ and $H^{2}(X,T_{X})=0$. \\
    Now $H^{i}(X,\wedge^{2}T_{X})=H^{i}(X,\mathcal{K}_{X}\otimes\wedge^{2}T_{X}\otimes\wedge^{2}T_{X})$.
 Since $\wedge^{2}T_{X}\otimes\wedge^{2}T_{X}$ is a positive line bundle, by Kodaira vanishing theorem,
 we have $H^{i}(X,\wedge^{2}T_{X})=0$,   $\forall i>0$.

\item Let $X=B_{r}(5\leq r\leq8)$.\\
If  $i>0, (i,j)\neq(1,1)$,  the assertion  can be proved in a similar way as in the case $1\leq r\leq4$.
When  $(i,j)=(1,1)$, the proof can be found in (\cite{kod},p226).
\end{enumerate}
 \qed

\section{Poisson Cohomology}
\begin{defi}
\label{LSX}
Let $(X, \pi)$ be a holomorphic Poisson manifold of dimension n.
 The Poisson cohomology  $H^\bullet_\pi (X) $ is
 the cohomology group of the complex of sheaves:
\begin{equation}
\mathcal{O}_{X}\xrightarrow{d_{\pi}}T_{X}\xrightarrow{d_{\pi}}.....
\xrightarrow{d_{\pi}}\wedge^{i-1}T_{X}\xrightarrow{d_{\pi}}\wedge^{i}T_{X}
\xrightarrow{d_{\pi}}\wedge^{i+1}T_{X}\xrightarrow{d_{\pi}}......
\xrightarrow{d_{\pi}}\wedge^{n}T_{X},
\end{equation}
where $d_{\pi}=[\pi,\cdot]$ is the Schouten bracket with $\pi$.
\end{defi}

\begin{lem}\label{LSX}
  \cite{LSX} 
  The Poisson cohomology of a holomorphic
  Poisson manifold $(X,\pi)$ is isomorphic to the  total cohomology of the double complex \\
$$\begin{array}{ccccccc}
......& &......& &......& & \\
d_{\pi}\big\uparrow & & d_{\pi}\big\uparrow & & d_{\pi}\big\uparrow & &  \\
\Omega^{0,0}(X,T^{2,0}X) & \xrightarrow{\bar{\partial}} &
 \Omega^{0,1}(X, T^{2,0}X) & \xrightarrow{\bar{\partial}} &
 \Omega^{0,2}(X, T^{2,0}X) &\xrightarrow{\bar{\partial}}   &
 ......\\
 d_{\pi}\big\uparrow & & d_{\pi}\big\uparrow & & d_{\pi}\big\uparrow & &  \\
\Omega^{0,0}(X,T^{1,0}X) & \xrightarrow{\bar{\partial}} &
 \Omega^{0,1}(X, T^{1,0}X) & \xrightarrow{\bar{\partial}} &
 \Omega^{0,2}(X,T^{1,0}X) &\xrightarrow{\bar{\partial}}   &
 ......\\
 d_{\pi}\big\uparrow & & d_{\pi}\big\uparrow & & d_{\pi}\big\uparrow & &  \\
\Omega^{0,0}(X, T^{0,0}X) & \xrightarrow{\bar{\partial}} &
\Omega^{0,1}(X, T^{0,0}X) & \xrightarrow{\bar{\partial}} & \Omega^{0,2}(X, T^{0,0}X) & \xrightarrow{\bar{\partial}} &
 ......\\
\end{array}$$
\\
\end{lem}

\begin{lem}\label{coho}
Let $(X,\pi)$ be a holomorphic Poisson manifold. If all the higher
cohomology groups $H^{i}(X,\wedge^{j}T_{X})$ vanish for $i>0$, then the
Poisson cohomology  $H^\bullet_\pi (X)$
is isomorphic to the cohomology of the complex
\begin{equation}
H^{0}(X,\mathcal{O}_{X})\xrightarrow {d_{\pi}}
H^{0}(X,T_{X})\xrightarrow{d_{\pi}}
H^{0}(X,\wedge^{2}T_{X})\xrightarrow{d_{\pi}}
H^{0}(X,\wedge^{3}T_{X})\xrightarrow{d_{\pi}}
\end{equation}
where $d_{\pi}=[\pi,\cdot]$.
\end{lem}

According to Theorem \ref{del}, the assumption in Lemma \ref{coho}
is satisfied   for all Del Pezzo Poisson surfaces except for  $B_{r}(5\leq r\leq8)$.

  When $X=\mathbb{CP}^{2}$, we have $H^{0}(X,\mathcal{O}_{X})=\mathbb{C},
  \quad H^{0}(X,\wedge^{i}T_{X})=0~~(i>2)$. According to Lemma \ref{coho}, the Poisson cohomology of  $(\mathbb{CP}^{2},\pi)$ can be expressed as follows:
  
     \begin{eqnarray*}
     \begin{cases}
 H^{0}_{\pi}(X)=\mathbb{C} \\
 H^{1}_{\pi}(X)=\text{Ker}\{d_{\pi}:H^{0}(X,T_{X}) \xrightarrow{d_{\pi}} H^{0}(X,\wedge^{2}T_{X})\}\\
 H^{2}_{\pi}(X)=\text{Coker}\{d_{\pi}:H^{0}(X,T_{X})\xrightarrow{d_{\pi}}H^{0}(X,\wedge^{2}T_{X})\}\\
 H^{i}_{\pi}(X)=0 ~~~(i>2),
 \end{cases}
     \end{eqnarray*}
     where, by Equation (\ref{cp2}), $d_{\pi}$ is a linear map from $\mathbb{C}^{8}$ 
     to $\mathbb{C}^{10}$, which is  denoted by $A_{\pi}$.

     On $U_{1}$, $\pi=h(x,w)\frac{\partial}
     {\partial x}\wedge \frac{\partial}{\partial w}$, $v=f(x,w)\frac{\partial}{\partial x}+ g(x,w)\frac{\partial}{\partial w}$,
     where $h,f,g$ are 
 as in Equations (\ref{cp2b})-(\ref{cp2v}). Then
     \begin{eqnarray*}
     & &[\pi,v]=[h(x,w)\frac{\partial}
     {\partial x}\wedge \frac{\partial}{\partial w},
     f(x,w)\frac{\partial}{\partial x}+ g(x,w)\frac{\partial}{\partial w}]\\
     & &=[(h\frac{\partial f}{\partial x}-f\frac{\partial h}{\partial x})+
     (h\frac{\partial g}{\partial w}-g\frac{\partial h}{\partial w})]
     \frac{\partial}{\partial x}\wedge \frac{\partial}{\partial w}.
     \end{eqnarray*}
     
     Let $\{\pi_{i} |i=1,2,...10\}$
     be  a basis in $H^{0}(X,\wedge^{2}T_{X})$ corresponding to
     $a_{k}=\delta_{ik}$, $k=1,2,...10$, in Equation (\ref{cp2b}), and
      $\{v_{i}|i=1,2,...8\}$ a basis  of $H^{0}(X,T_{X})$
      corresponding to $b_{k}=\delta_{ik}$, $k=1,2,...,8 $ in Equation (\ref{cp2v}).
       Under these bases,
     the linear map $A_{\pi}$ can be written as the following $10\times 8$ matrix: \\
      \begin{equation}\label{mcp2}
    A_{\pi}=\left(\begin{array}{cccccccc}
    -a_{2} & a_{1} & 0 & -a_{3}& 0& a_{1}& 0& 0\\
    -2a_{4}& 0& 0& -a_{5}&-a_{3}& a_{2}&3a_{1}& 0\\
    -a_{5} & a_{3}& -a_{2}& -2a_{6}& 0& 0& 0&3a_{1}\\
    -3a_{7}&-a_{4}& 0 &-a_{8}&-a_{5}&a_{4}&2a_{2}&0\\
    -2a_{8}& 0&-2a_{4}&-2a_{9}&-2a_{6}&0&2a_{3}&2a_{2}\\
    -a_{9}& a_{6}&-a_{5}&-3a_{10}& 0&-a_{6}& 0&2a_{3}\\
      0 &-2a_{7}& 0 & 0 &-a_{8}& a_{7} &a_{4}& 0\\
      0& -a_{8}&-3a_{7}& 0 & -2a_{9}& 0&a_{5}&a_{4}\\
      0 & 0 & -2a_{8}& 0 &-3a_{10}& -a_{9}&a_{6}&a_{5}\\
      0 &a_{10}&-a_{9}& 0& 0&-2a_{10}& 0&a_{6}\\
    \end{array}\right)
    \end{equation}

    \begin{thm}
    Let $X=\mathbb{CP}^{2}$ and $\pi$ a holomorphic bivector field of the form as in Equation (\ref{cp2b}). Then the Poisson cohomology of $(X, \pi)$ is given as follows.
     \begin{eqnarray*}
     \begin{cases}
     \dim H^{0}_{\pi}(X)=1\\
     \dim H^{1}_{\pi}(X)=8-rank(A_{\pi})\\
      \dim H^{2}_{\pi}(X)=10-rank(A_{\pi})\\
     \dim H^{i}_{\pi}(X)=0 \quad (i>2),\\
      \end{cases}
     \end{eqnarray*}
     where $A_{\pi}$ is the matrix as in Equation (\ref{mcp2}).
     \end{thm}
    \begin{ex}
    \begin{enumerate}
    \item If $\pi=\frac{\partial}{\partial x}\wedge \frac{\partial}{\partial w}$,
    then rank $(A_{\pi})=3$ and therefore we have
    $$\dim H^{1}_{\pi}(\mathbb{CP}^{2})=5,\quad \dim H^{2}_{\pi}(\mathbb{CP}^{2})=7.$$
    \item If $\pi=x\frac{\partial}{\partial x}\wedge \frac{\partial}{\partial w}$,
     then rank$(A_{\pi})=5$ and therefore
     $$\dim H^{1}_{\pi}(\mathbb{CP}^{2})=3,\quad \dim H^{2}_{\pi}(\mathbb{CP}^{2})=5.$$
    \item If $\pi= xw\frac{\partial}{\partial x}\wedge \frac{\partial}{\partial w}$,
    then rank$(A_{\pi})=6$ and therefore
     $$\dim H^{1}_{\pi}(\mathbb{CP}^{2})=2,~~~~\dim H^{2}_{\pi}(\mathbb{CP}^{2})=4.$$
     \end{enumerate}
    \end{ex}

    Similarly, we can prove the following

     \begin{thm}
     Let $X=\mathbb{CP}^{1}\times \mathbb{CP}^{1}$ and $\pi$ a holomorphic bivector field  as in Equation (\ref{cp1b}). The Poisson cohomology of $(X, \pi)$  is then given as follows.
     \begin{align*}
     \dim H^{0}_{\pi}(X) &=1, \\
     \dim H^{1}_{\pi}(X) &=6-rank(A_{\pi}), \\
     \dim H^{2}_{\pi}(X) &=9-rank(A_{\pi}), \\
     \dim H^{i}_{\pi}(X) &=0 \quad (i>2),
     \end{align*}
     where $A_{\pi}$ is the matrix
     \begin{equation}\label{mcp1}
    A_{\pi}=\left(\begin{array}{cccccc}
    -a_{2} & a_{1} & 0 & -a_{4}&  a_{1}& 0\\
    -2a_{3}& 0  & 2a_{1} & -a_{5} &a_{2}   & 0\\
     0  & -a_{3}& a_{2} & -a_{6}&a_{3}& 0 \\
     -a_{5}&a_{4}& 0&-2a_{7}& 0& 2a_{1}\\
     -2a_{6}& 0&2a_{4}&-2a_{8}& 0&2a_{2}\\
     0 &-a_{6}&a_{5}&-2a_{9}& 0&2a_{3}\\
     -a_{8} &a_{7}& 0& 0&-a_{7}&a_{4}\\
     -2a_{9}& 0&2a_{7}& 0& -a_{8}&a_{5}\\
     0 &-a_{9}&a_{8}& 0&-a_{9}&a_{6}\\
    \end{array}\right)
    \end{equation}
     \end{thm}

     \begin{ex}
     \begin{enumerate}
     \item In the case $\pi=\frac{\partial}{\partial x}\wedge \frac{\partial}{\partial w}$,
     since rank$(A_{\pi})=3$, we have
     $$\dim H^{1}_{\pi}(\mathbb{CP}^{1}\times \mathbb{CP}^{1})=3,\quad \dim H^{2}_{\pi}(\mathbb{CP}^{1}\times \mathbb{CP}^{1})=6.$$
    \item In the case $\pi=x\frac{\partial}{\partial x}\wedge \frac{\partial}{\partial w}$,
    since rank$(A_{\pi})=4$, we have
    $$\dim H^{1}_{\pi}(\mathbb{CP}^{1}\times \mathbb{CP}^{1})=2,\quad\dim H^{2}_{\pi}(\mathbb{CP}^{1}\times \mathbb{CP}^{1})=5.$$
    \end{enumerate}
    \end{ex}

 Consider a Poisson Del Pezzo surface $B_{r} (1\leq r\leq8)$. By Lemma (\ref{kodb}),
 we have the following commutative diagram:
 \begin{equation}
 \begin{array}{ccc}
 H^{0}(B_{r},T_{B_{r}}) &\xrightarrow{d_{\pi}}& H^{0}(B_{r},\wedge^{2}T_{B_{r}})\\
 \mathrm{\rho}_{*}\big\updownarrow & &\mathrm{\rho}_{*}\big\updownarrow\\
 V^{1}_{r} & \xrightarrow{d_{\mathrm{\rho}_{*}\pi}} & V^{2}_{r}
 \end{array}
 \end{equation}
 As a consequence, we have

 \begin{thm}
 For $(B_{r},\pi)$ $(1\leq r\leq4)$, the Poisson cohomology is given as follows.
   \begin{eqnarray}
   \begin{cases}
 \dim H^{0}_{\pi}(B_{r})=1 \\
 \dim H^{1}_{\pi}(B_{r})=8-2r-rank(A_{\pi})\\
 \dim H^{2}_{\pi}(B_{r})=10-r-rank(A_{\pi})\\
 \dim H^{i}_{\pi}(B_{r})=0 ~~~ (i>2),
 \end{cases}
 \end{eqnarray}
 where $A_{\pi}$ is the matrix defined as follows.
 \begin{enumerate}
 \item If $r=1$, on $U_{0}=\{[z_{0},z_{1},z_{2}]|z_{0}\neq 0\}$, $\mathrm{\rho}_{*}\pi$ can be written as
 \begin{equation*}
 \mathrm{\rho}_{*}\pi=(a_{2}x+a_{3}w+a_{4}x^{2}+a_{5}xw+a_{6}w^{2}+a_{7}x^{3}+a_{8}x^{2}w+
 a_{9}xw^{2}+a_{10}w^{3})\frac{\partial}{\partial x}\wedge\frac{\partial}{\partial w}.
  \end{equation*}
  Then $A_{\pi}$ has the following form:
 $$A_{\pi}=\left(\begin{array}{cccccc}
     0 & 0 & -a_{3} & a_{2}& 0 & 0\\
     a_{3} & -a_{2} & 0 & 0 & 0 & 0\\
     -a_{4}  & 0 & -a_{5} & a_{4}& 2a_{2}& 0 \\
     0 &-2a_{4}& -2a_{6}& 0& 2a_{3}& 2a_{2}\\
     a_{6}& -a_{5}& 0 &-a_{6}& 0&2a_{3}\\
     -2a_{7} &0 &-a_{8}&a_{7}& a_{4}& 0\\
     -a_{8} &-3a_{7}& -2a_{9}& 0&a_{5}&a_{4}\\
     0 & -2a_{8}&-3a_{10}& -a_{9}& a_{6}&a_{5}\\
     a_{10} &-a_{9}&0 & -2a_{10}& 0&a_{6}\\
    \end{array}\right)$$
 \item If $r=2$, on $U_{0}=\{[z_{0},z_{1},z_{2}]|z_{0}\neq 0\}$, $\mathrm{\rho}_{*}\pi$ can be written as:
 \begin{equation*}
 \mathrm{\rho}_{*}\pi=(a_{2}x+a_{3}w+a_{4}x^{2}+a_{5}xw+a_{6}w^{2}-(a_{2}+a_{4})x^{3}+a_{8}x^{2}w+
 a_{9}xw^{2}+a_{10}w^{3})\frac{\partial}{\partial x}\wedge\frac{\partial}{\partial w}.
  \end{equation*}
  Then $A_{\pi}$ has the following form:
 $$A_{\pi}=\left(\begin{array}{cccc}
     0 & 0 &  a_{2}& 0\\
    a_{3}& -a_{2}  & 0  & 0\\
     -2a_{2}-a_{4} & 0 &a_{4}& 0 \\
     -2a_{3}&-2a_{4}& 0& 2a_{2}\\
     a_{6}& -a_{5}&-a_{6} &2a_{3}\\
     -a_{8}-a_{5} &3a_{2}+3a_{4}& 0& a_{4}\\
     -a_{6}& -2a_{8}& -a_{9}&a_{5}\\
     a_{10}&-a_{9}& -2a_{10}&a_{6}\\
    \end{array}\right)$$
 \item If $r=3$, on $U_{0}=\{[z_{0},z_{1},z_{2}]|z_{0}\neq 0\}$, $\mathrm{\rho}_{*}\pi$ can be written as:
\begin{equation*}
 \mathrm{\rho}_{*}\pi=(a_{2}x+a_{3}w+a_{4}x^{2}+a_{5}xw+a_{6}w^{2}-(a_{2}+a_{4})x^{3}+a_{8}x^{2}w+
 a_{9}xw^{2}-(a_{3}+a_{6})w^{3})\frac{\partial}{\partial x}\wedge\frac{\partial}{\partial w}.
  \end{equation*}
 Then $A_{\pi}$ has the following form:
 $$A_{\pi}=\left(\begin{array}{cc}
     0 & a_{2} \\
     a_{3}& 0  \\
     -2a_{2}-a_{4} & a_{4}\\
     -2a_{3}&-2a_{2}\\
     a_{6}& -2a_{3}-a_{6}\\
     -a_{5}-a_{8} &-a_{4}\\
     -a_{6}& -a_{5}-a_{9}\\
    \end{array}\right)$$
 \item If $r=4$, then $A_{\pi}=0$ and therefore
 \begin{eqnarray}
 \begin{cases}
 \dim H^{0}_{\pi}(B_{4})=1 \\
 \dim H^{1}_{\pi}(B_{4})=0\\
 \dim H^{2}_{\pi}(B_{4})=6\\
 \dim H^{i}_{\pi}(B_{4})=0 & (i>2)
 \end{cases}
 \end{eqnarray}
 \end{enumerate}
 \end{thm}
\proof By $A_{\pi}$, we denote the linear map $V^{1}_{r}\xrightarrow{d_{\mathrm{\rho}_{*}\pi}} V^{2}_{r}$.
 \begin{enumerate}
 \item Consider the case $X=B_{1}$, which is $\mathbb{CP}^{2}$ blow up at $p_{1}=[1,0,0]$.
  By Lemma (\ref{kodb}), and Equations (\ref{cp2b})-(\ref{cp2v}), 
  on $U_{0}=\{[z_{0},z_{1},z_{2}]|z_{0}\neq 0\}$, any $\pi\in V^{2}_{1}$ and
 $v\in V^{1}_{1}$ must be of the following forms:
 \begin{equation}
 \pi=(a_{2}x+a_{3}w+a_{4}x^{2}+a_{5}xw+a_{6}w^{2}+a_{7}x^{3}+a_{8}x^{2}w+
 a_{9}xw^{2}+a_{10}w^{3})\frac{\partial}{\partial x}\wedge\frac{\partial}{\partial w},
  \end{equation}
 \begin{equation}
 v=(b_{2}x+b_{3}w+b_{7}x^{2}+b_{8}xw)\frac{\partial}{\partial x}+
 (b_{5}x+b_{6}w+b_{7}xw+b_{8}w^{2})\frac{\partial}{\partial w}.
 \end{equation}
 Let $\pi_{i}\in V^{2}_{1}$, $i=2,...10$,  be a basis of $V^{2}_{1}$ corresponding to $a_{k}=\delta_{ik}$, $k=2,...10 $;
  $v_{i}\in V^{1}_{1}$, $i=2,3,5,6,7,8$ be a basis corresponding to $b_{k}=\delta_{ik}$, $k=2,3,5,6,7,8 $.
 Under these bases, the linear map $A_{\pi}$ can be expressed by the following $9\times 6$ matrix:
 $$A_{\pi}=\left(\begin{array}{cccccc}
     0 & 0 & -a_{3} & a_{2}& 0 & 0\\
     a_{3} & -a_{2} & 0 & 0 & 0 & 0\\
     -a_{4}  & 0 & -a_{5} & a_{4}& 2a_{2}& 0 \\
     0 &-2a_{4}& -2a_{6}& 0& 2a_{3}& 2a_{2}\\
     a_{6}& -a_{5}& 0 &-a_{6}& 0&2a_{3}\\
     -2a_{7} &0 &-a_{8}&a_{7}& a_{4}& 0\\
     -a_{8} &-3a_{7}& -2a_{9}& 0&a_{5}&a_{4}\\
     0 & -2a_{8}&-3a_{10}& -a_{9}& a_{6}&a_{5}\\
     a_{10} &-a_{9}&0 & -2a_{10}& 0&a_{6}\\
    \end{array}\right)$$

  \item Consider the case $X=B_{2}$, which is $\mathbb{CP}^{2}$ blow up at $p_{1}=[1,0,0]$ and $p_{2}=[1,1,0]$.
   By Lemma \ref{kodb}, and Equations (\ref{cp2b}) (\ref{cp2v}), any $\pi\in V^{2}_{2}$
   and $v\in V^{1}_{2}$ must be of the following forms on $U_{0}=\{[z_{0},z_{1},z_{2}]|z_{0}\neq 0\}$:
 \begin{equation}
 \pi=(a_{2}x+a_{3}w+a_{4}x^{2}+a_{5}xw+a_{6}w^{2}-(a_{2}+a_{4})x^{3}+a_{8}x^{2}w+
 a_{9}xw^{2}+a_{10}w^{3})\frac{\partial}{\partial x}\wedge\frac{\partial}{\partial w},
  \end{equation}
 \begin{equation}
 v=(b_{2}x+b_{3}w-b_{2}x^{2}+b_{8}xw)\frac{\partial}{\partial x}+
 (b_{6}w-b_{2}xw+b_{8}w^{2})\frac{\partial}{\partial w}.
 \end{equation}
  Hence $A_{\pi}$ can be written as the following $8\times 4$ matrix:
 $$A_{\pi}=\left(\begin{array}{cccc}
     0 & 0 &  a_{2}& 0\\
    a_{3}& -a_{2}  & 0  & 0\\
     -2a_{2}-a_{4} & 0 &a_{4}& 0 \\
     -2a_{3}&-2a_{4}& 0& 2a_{2}\\
     a_{6}& -a_{5}&-a_{6} &2a_{3}\\
     -a_{8}-a_{5} &3a_{2}+3a_{4}& 0& a_{4}\\
     -a_{6}& -2a_{8}& -a_{9}&a_{5}\\
     a_{10}&-a_{9}& -2a_{10}&a_{6}\\
    \end{array}\right)$$

 \item Consider the case $X=B_{3}$, which is $\mathbb{CP}^{2}$ blow up at $p_{1}=[1,0,0],p_{2}=[1,1,0]$ and $p_{3}=[1,0,1]$. By Lemma (\ref{kodb}), and Equations (\ref{cp2b})-(\ref{cp2v}), any $\pi
 \in V^{2}_{2}$ and $v\in V^{1}_{2}$ are of the following form on $U_{0}=\{[z_{0},z_{1},z_{2}]|z_{0}\neq 0\}$:
 \begin{equation}
 \pi=(a_{2}x+a_{3}w+a_{4}x^{2}+a_{5}xw+a_{6}w^{2}-(a_{2}+a_{4})x^{3}+a_{8}x^{2}w+
 a_{9}xw^{2}-(a_{3}+a_{6})w^{3})\frac{\partial}{\partial x}\wedge\frac{\partial}{\partial w},
  \end{equation}
 \begin{equation}
 v=(b_{2}x-b_{2}x^{2}-b_{6}xw)\frac{\partial}{\partial x}+
 (b_{6}w-b_{2}xw-b_{6}w^{2})\frac{\partial}{\partial w}.
 \end{equation}
 Hence $A_{\pi}$ can be written as the following $7\times2$ matrix:
 $$A_{\pi}=\left(\begin{array}{cc}
     0 & a_{2} \\
     a_{3}& 0  \\
     -2a_{2}-a_{4} & a_{4}\\
     -2a_{3}&-2a_{2}\\
     a_{6}& -2a_{3}-a_{6}\\
     -a_{5}-a_{8} &-a_{4}\\
     -a_{6}& -a_{5}-a_{9}\\
    \end{array}\right)$$

 \item Consider the case $X=B_{4}$. According to Equation (\ref{bhv}), $H^{0}(X,T_{X})=0$. Hence $A_{\pi}=0$,
 and therefore
 $$\dim H^{1}_{\pi}(B_{4})=0,~~~\dim H^{2}_{\pi}(B_{4})=6.$$
 \end{enumerate}

 \qed

\begin{thm}
 For $(B_{r},\pi)(5\leq r\leq8)$, the Poisson cohomology is
   \begin{eqnarray}
   \begin{cases}
 \dim H^{0}_{\pi}(B_{r})=1 \\
 \dim H^{1}_{\pi}(B_{r})=0\\
 \dim H^{2}_{\pi}(B_{r})=r+2\\
 \dim H^{i}_{\pi}(B_{r})=0 ~~~(i>2).
 \end{cases}
 \end{eqnarray}
 \end{thm}
\proof We will use spectral sequence associated to the double complex in Lemma (\ref{LSX}).
 For the double complex $$K=\underset{0\leq i,j\leq n}{\bigoplus}K^{i,j},$$
 where $K^{i,j}=\Omega^{0,i}(B_{r},T^{j,0}B_{r})$
  and the differential $D=d_{\pi}+\overline{\partial}$.
 Denote $H^{i,j}_{\overline{\partial}}(K)=H^{i}(B_{r},\wedge^{j}T_{B_{r}})$. By Theorem (\ref{del}) and Equation (\ref{bhv}),
 $$\dim H^{0,0}_{\overline{\partial}}(K)=1,~~~\dim H^{0,2}_{\overline{\partial}}(K)=10-r,
 ~~~\dim H^{1,1}_{\overline{\partial}}(K)=2r-8,$$
  and $H^{i,j}_{\overline{\partial}}(K)=0$, for all other $i,j$.\\
  Since $$E^{i,j}_{1}=H^{i,j}_{\overline{\partial}}(K)$$
  and $$E_{1}=E_{2}=...E_{\infty}, ~~~ d_{1}=d_{2}=...=0,$$
  the total cohomology of the double complex is
  $$H^{0}_{\pi}(B_{r})=E^{0,0}_{1},~~~H^{2}_{\pi}(B_{r})=\underset{i+j=2}\oplus E^{i,j}_{\infty}=E^{0,2}_{1}\oplus E^{1,1}_{1},$$
  and $$H^{k}_{\pi}(B_{r})=\underset{i+j=k}\oplus E^{i,j}_{\infty}=0,$$ if $k\neq0,2$.
  By Theorem (\ref{del}) and Equation (\ref{bhb}), we have
  \begin{eqnarray*}
   \begin{cases}
 \dim H^{0}_{\pi}(B_{r})=1 \\
 \dim H^{1}_{\pi}(B_{r})=0\\
 \dim H^{2}_{\pi}(B_{r})=(10-r)+(2r-8)=r+2\\
 \dim H^{i}_{\pi}(B_{r})=0 ~~~ (i>2).
 \end{cases}
 \end{eqnarray*}

\qed

\end{document}